\documentclass[aps,showpacs,epsfig]{revtex4}
\usepackage{graphicx}
\usepackage{amsmath}
\usepackage{array}
\begin{document}
\title{Quantum random walk of two photons in separable and entangled state}
\author{P. K. Pathak\footnote{On leave from
Physical Research Laboratory, Navrangpura, Ahmedabad-380 009, India}
and G. S. Agarwal}
\affiliation{Department of Physics, Oklahoma
State University, Stillwater, Oklahoma 74078, USA}
\date{\today}
\begin{abstract}
We discuss quantum random walk of two photons using linear optical
elements. We analyze the quantum random walk using photons in a
variety of quantum states including entangled states. We find that
for photons initially in separable Fock states, the final state is
entangled. For polarization entangled photons produced by type II
downconverter, we calculate the joint probability of detecting two
photons at a given site. We show the remarkable dependence of the
two photon detection probability on the quantum nature of the state.
In order to understand the quantum random walk, we present exact
analytical results for small number of steps like five. We present
in details numerical results for a number of cases and supplement
the numerical results with asymptotic analytical results.
\end{abstract}
\maketitle
\section{Introduction}
A new paradigm in the study of random walks has recently emerged
\cite{aharonov,kempe, sanders, knight,zhao, jeong, our, milburn,
lattice, exp, bouwmeester, bose1,bose2,zubairy, chaos, nmr,
ent-knight}. The random walker has been assigned an additional
quantum degree of freedom which could be spin degree of freedom
\cite{aharonov}. Thus walker goes left or right depending on the
spin degree of freedom. The probability of finding the walker at a
given site now depends on the spin state of the walker. All this has
now been well studied \cite{kempe} and numerical simulations have
been done to find the site distribution after walker has taken large
number of steps. Using linear optical elements, Do et al \cite{exp}
have realized the quantum random walk (QRW). However, in their
experiments they used a weak coherent field rather than a field with
strong quantum character. This is fine if we recall the results of
Knight et al \cite{knight} who showed that QRW of a single walker
can be implemented by using classical fields. A similar arrangement
is discussed earlier by other authors as well \cite{zhao,jeong}.
Jeong et al \cite{jeong} analyzed the cases of a walker in a
coherent state and in a single photon state and concluded that the
final probability distribution was identical in the two cases;
though different from that of the classical random walk (i.e. a
walker without the additional quantum degree of freedom). This is
explained by Jeong et al \cite{jeong} in terms of the
P-representation of the state of photons. Thus an important question
is-what would be a strict QRW which can not be produced by using
classical fields. To understand this aspect, we study QRW by two
photons in a variety of quantum states including entangled states.
We find that even if initially the two photons are in separable Fock
states, the final state is entangled. This is quite an interesting
quantum property and has no classical counter part. Omar et al
\cite{bose2} have studied QRW of two nonidentical walkers with
entangled initial state and have shown that the final state is
entangled depending on the initial entanglement. Our model of QRW is
different from that of Ref \cite{bose2} as our system can produce
entanglement even if initially there is none. This is clearly borne
out by our result in Sec.IV A. Further, we calculate photon-photon
correlations which in the past have been very successfully used to
reveal the quantum character of the fields \cite{mandel,kwiat}. We
show the remarkable dependence of the two photon detection
probability on the quantum nature of the state of photons. We
present explicit results for all four Bell states of the incoming
photons. Our work thus brings out the role of coherences and
entanglement in QRW.

The organization of the paper is as follows. In Sec.II, we present
the system of linear optical elements used to realize QRW. In
Sec.III, we discuss QRW of a single photon when the photon enters in
the arrangement through the two input ports. The state of the
incoming photon is a pure state. We present an approximate analysis
for the probability of finding the photon at a site after a large
number of steps \cite{nayak,brun} and compare with the exact
numerical results. In Secs.IV and V, we consider QRW of two photons.
We derive analytical results for the photon-photon entanglement
corresponding to the case when the walkers take only a small number
of steps. The analytical results clearly demonstrate the dependence
of two photon detection probability on entanglement between the
walkers. We evolve a numerical strategy as well as an approximate
analysis to obtain results for final state of the photons after a
large number of steps. In Sec.IV, we consider the case when the
initial state of the photons is separable however the final state of
the photons is entangled. Here the entanglement is generated due to
the passage of photons through linear optical elements. We also
discuss a case when two photons in Fock state are replaced by two
photons in coherent states. We point out that QRW in our scheme with
photons either in separable Fock states or in an entangled state can
not be reproduced by using coherent states. In Sec.V, we discuss QRW
of two photons when initial state is an entangled state and discuss
the dependence of two photon detection probability on entanglement
initially present in the state of the walkers. We present our
conclusions in Sec.VI.
\section{Description of the Arrangement}
\begin{figure}
\centering
\includegraphics[width=3in]{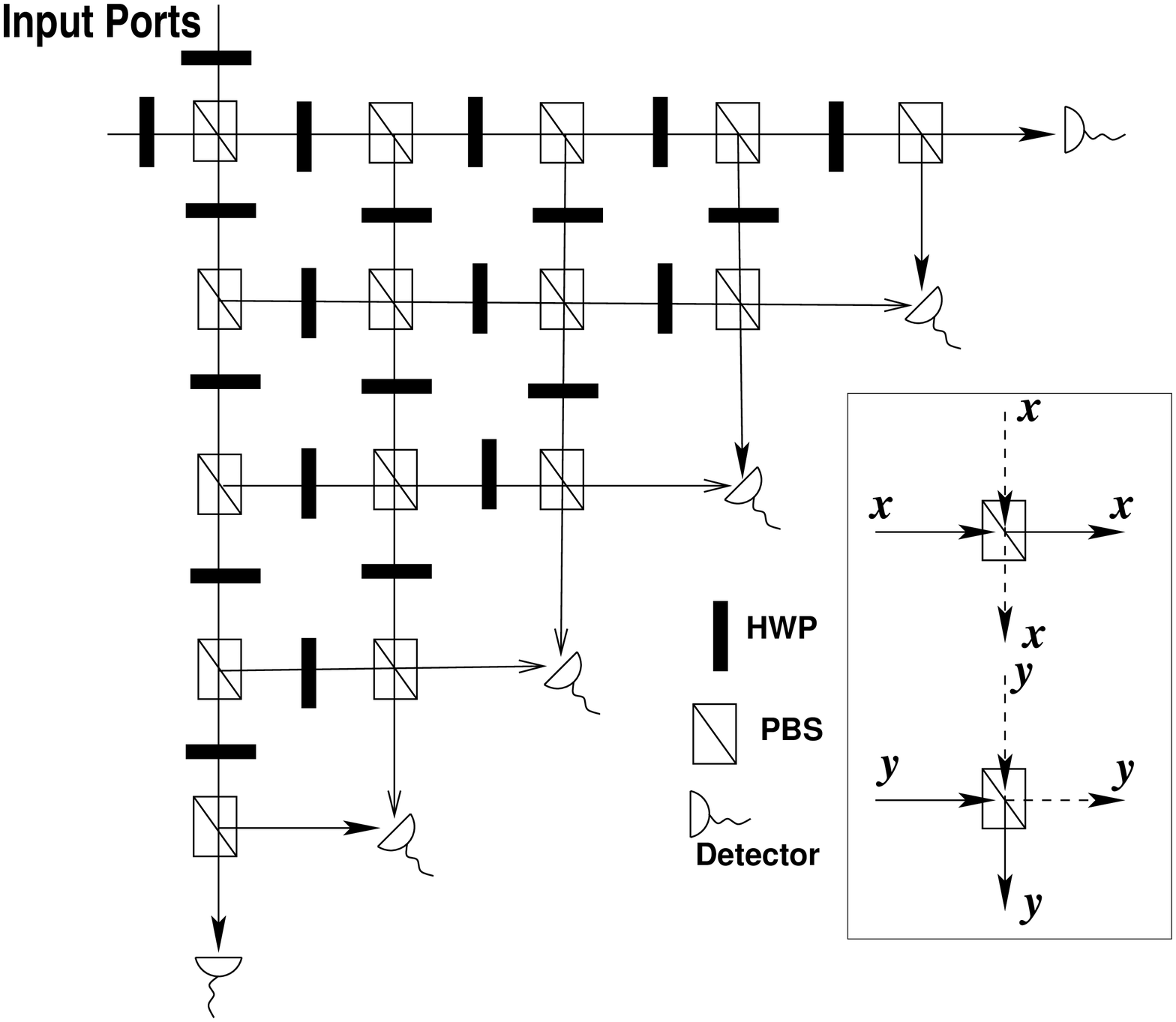}
\caption{The schematic arrangements for realizing QRW of two
entangled walkers. In the inset effect of polarization beam splitter
is shown explicitly. The x-polarized photon is transmitted in the
same direction but the y-polarized photon is reflected and changed
the direction of propagation.} \label{fig1}
\end{figure}

In Fig. \ref{fig1}, we show a schematic arrangement for realization
of QRW on a line using linear optical elements like polarizing beam
splitters (PBS) and half wave plates (HWP). The arrangement looks
like a large interferometer and has been discussed in an
experimental realization of quantum quincunx \cite{exp}. The photons
enter in the arrangement through the two input ports. The state of
the photons anywhere along the arrangement can be completely defined
in terms of its direction of propagation and state of polarization.
A half wave plate performs Hadamard operation for a single photon in
polarization state basis \cite{hwp},
\begin{eqnarray}
&&|x\rangle\xrightarrow{HWP}\frac{1}{\sqrt{2}}\left(|x\rangle+|y\rangle\right),\nonumber\\
&&|y\rangle\xrightarrow{HWP}\frac{1}{\sqrt{2}}\left(|x\rangle-|y\rangle\right),
\label{hadamard}
\end{eqnarray}
where $|x\rangle$ and $|y\rangle$ are two orthogonal linear
polarization states of the photons. The polarizing beam splitter
(PBS) produces polarization dependent spatial displacement in the
position of the photons. Clearly the displacement will be either in
horizontal or in vertical direction depending on the direction of
propagation and the state of polarization of the incoming photons.
The operation of PBS can be written as
\begin{equation}
\left(\begin{array}{c}|h,x\rangle\\|h,y\rangle\\|v,x\rangle\\|v,y\rangle\end{array}\right)
\xrightarrow{PBS}\left(\begin{array}{cccc}1&0&0&0\\0&0&0&1\\0&0&1&0
\\0&1&0&0\end{array}\right)\left(\begin{array}{c}|h,x\rangle\\
|h,y\rangle\\
|v,x\rangle\\
|v,y\rangle\end{array}\right), \label{disp}
\end{equation}
where $h(v)$ shows the propagation of the photons along the
horizontal (vertical) direction. After passing through PBS photons
are displaced by one step in the horizontal direction when they come
in the state $|h,x\rangle$ or $|v,y\rangle$ and displaced by one
step in vertical direction when they come in the state $|h,y\rangle$
or $|v,x\rangle$. Thus, in our scheme, states $|h,x\rangle$,
$|h,y\rangle$, $|v,x\rangle$, and $|v,y\rangle$ are corresponding to
the states of a four sided coin and the position of the photon along
the optical arrangement performs QRW.

The photons enter through the input ports and pass through a half
wave plate followed by a PBS. The half wave plate performs Hadamard
operation in the polarization space. Then PBS produces shift in the
position of the photons depending on the polarization and the
direction of the propagation of the incoming photons. Thus each
layer of half wave plate followed by the PBS generates one step of
QRW in the spatial state of the photons. After particular number of
such steps the position of the photons is detected by an array of
the detectors. The detection of the photons can be made polarization
sensitive or insensitive depending on the particular interest of
measurement of the state.

In Fig. \ref{fig1}, we show the arrangement which has five layers,
each layer is made of half wave plate followed by a PBS. This
arrangement can produce QRW by five steps. Needless to say that for
large number of steps a large number of optical elements will be
required and such a big arrangement may have some difficulties in
alignment as well as the visibility of the signal.
\section{QRW of a single photon}
In this section, we consider the case when a single photon enters
through the input ports. Thus we have a single walker and the QRW is
controlled by a four sided coin. The four sides of the coin are
formed by the four degrees of freedom of the photon, two directions
of propagation as the photon can move in horizontal or in the
vertical direction, and two directions of polarization. The state of
the photon at any time during the QRW can be defined as
$|n,q,j,k\rangle$, where $n$ represents the number of steps, $q$
represents the spatial position of the photon with the convention
that the horizontal displacement is in $+ve$ direction and the
vertical displacement is in $-ve$ direction, and $j$, $k$ represent
the direction of propagation and the direction of polarization
respectively.

The dynamics of the photon can be described by expressing the state
of the photon after $n+1$ steps in terms of the states after $n$
steps in the following way.
\begin{eqnarray}
|n,q,h,x\rangle&&\xrightarrow{HWP}\frac{1}{\sqrt{2}}
(|n,q,h,x\rangle+|n,q,h,y\rangle)\nonumber\\
&&\xrightarrow{PBS}\frac{1}{\sqrt{2}}
(|n+1,q+1,h,x\rangle+|n+1,q-1,v,y\rangle),\label{it1}\\
|n,q,h,y\rangle&&\xrightarrow{HWP}\frac{1}{\sqrt{2}}
(|n,q,h,x\rangle-|n,q,h,y\rangle)\nonumber\\
&&\xrightarrow{PBS}\frac{1}{\sqrt{2}}
(|n+1,q+1,h,x\rangle-|n+1,q-1,v,y\rangle),\label{it2}\\
|n,q,v,x\rangle&&\xrightarrow{HWP}\frac{1}{\sqrt{2}}
(|n,q,v,x\rangle+|n,q,v,y\rangle)\nonumber\\
&&\xrightarrow{PBS}\frac{1}{\sqrt{2}}
(|n+1,q-1,v,x\rangle+|n+1,q+1,h,y\rangle),\label{it3}\\
|n,q,v,y\rangle&&\xrightarrow{HWP}\frac{1}{\sqrt{2}}
(|n,q,v,x\rangle-|n,q,v,y\rangle)\nonumber\\
&&\xrightarrow{PBS}\frac{1}{\sqrt{2}}(|n+1,q-1,v,x\rangle-|n+1,q+1,h,y\rangle).
\label{it4}
\end{eqnarray}
First the half wave plate generates transformation in the
polarization states of the photon which is followed by the spatial
transformation generated by the PBS. We can reexpress Eqs.
(\ref{it1})-(\ref{it4}) in terms of operators $C$ and $S$,
equivalently, as
\begin{equation}
\left[
\begin{array}{c}
|n+1,q+1,h,x\rangle\\
|n+1,q+1,h,y\rangle\\
|n+1,q-1,v,x\rangle\\
|n+1,q-1,v,y\rangle
\end{array}
\right]\equiv S\otimes C\left[
\begin{array}{c}
|n,q,h,x\rangle\\
|n,q,h,y\rangle\\
|n,q,v,x\rangle\\
|n,q,v,y\rangle\end{array} \right],
\label{evol}
\end{equation}
where $C$ and $S$ are
\begin{eqnarray}
C=\frac{1}{\sqrt{2}}\left(|h,x\rangle\langle h,x|+|h,x\rangle\langle
h,y|+|h,y\rangle\langle v,x|-|h,y\rangle\langle
v,y|\right.\nonumber\\
\left.+|v,x\rangle\langle v,x|+|v,x\rangle\langle
v,y|+|v,y\rangle\langle h,x|-|v,y\rangle\langle h,y|\right)~
\label{coin}\end{eqnarray}
\begin{eqnarray}
S\equiv |h,x\rangle\langle h,x|\otimes|q+1\rangle\langle
q|+|h,y\rangle\langle h,y|\otimes|q+1\rangle\langle q|\nonumber\\
+|v,x\rangle\langle v,x|\otimes|q-1\rangle\langle q|
+|v,y\rangle\langle v,y|\otimes|q-1\rangle\langle q| \label{shift}
\end{eqnarray}
Thus, the unitary coin operator $C$ and the conditioned shift
operator $S$ for the QRW of the photon are given by Eqs.
(\ref{coin}) and (\ref{shift}) respectively. The transformation
$U_s\equiv S\otimes C$ acting on the initial state of the photon is
equivalent to the one step of QRW. The iterative application of the
transformation $U_s$ for $n$ times, $U_s^n$, gives the QRW of $n$
steps.
\begin{figure}
\centering
\includegraphics[width=3in]{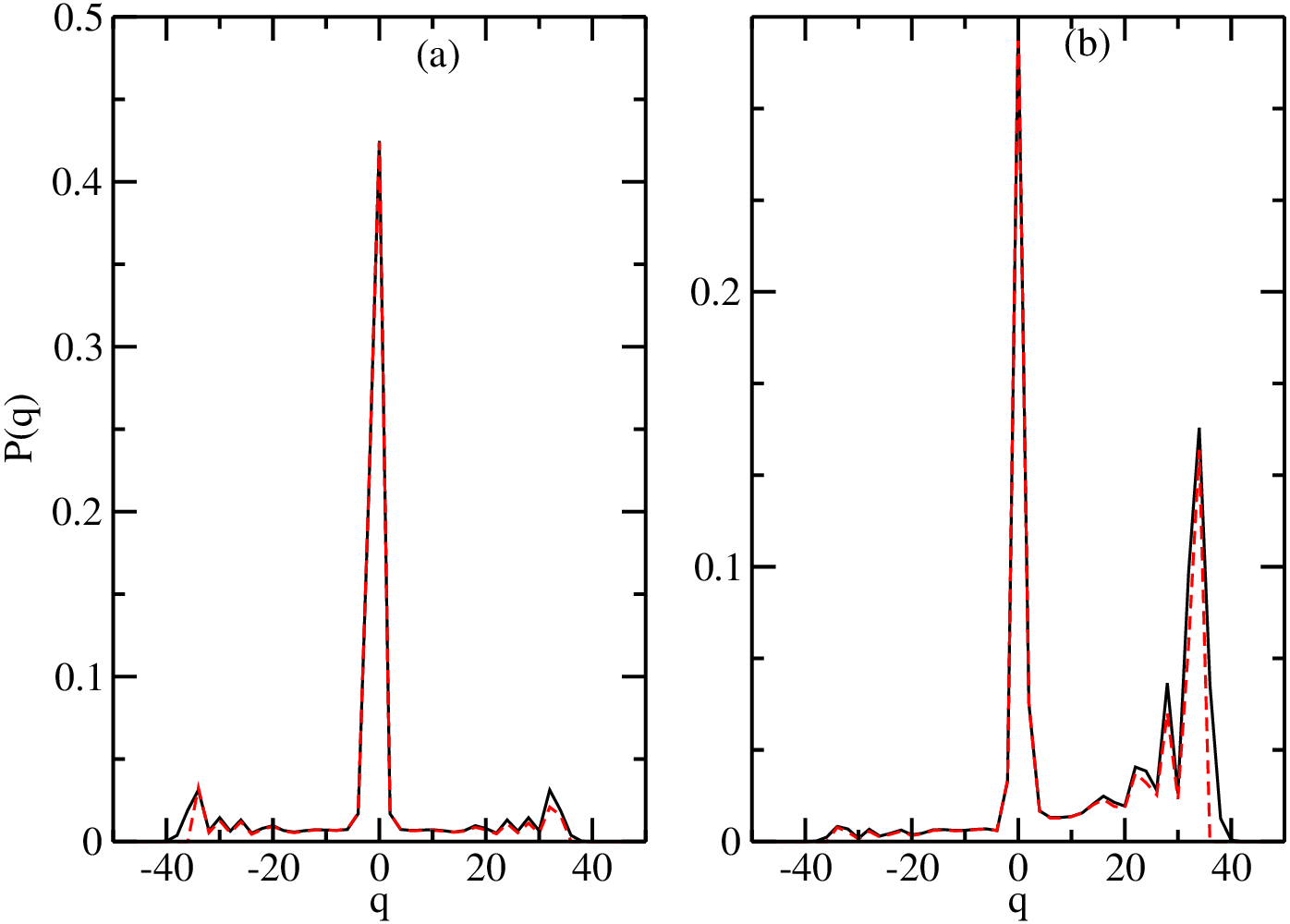}
\caption{(Color online) The probability distribution $P(q)$ for
detecting the walker at position $q$, for the initial coin state (a)
$\frac{1}{\sqrt{2}}(|h,x\rangle+|v,y\rangle)$ and
(b)$\frac{1}{\sqrt{2}}(|h,x\rangle-|v,y\rangle)$. The black solid
(red dashed) line is corresponding to exact (approximated) values.}
\label{fig2}
\end{figure}
\begin{figure}
\centering
\includegraphics[width=3in]{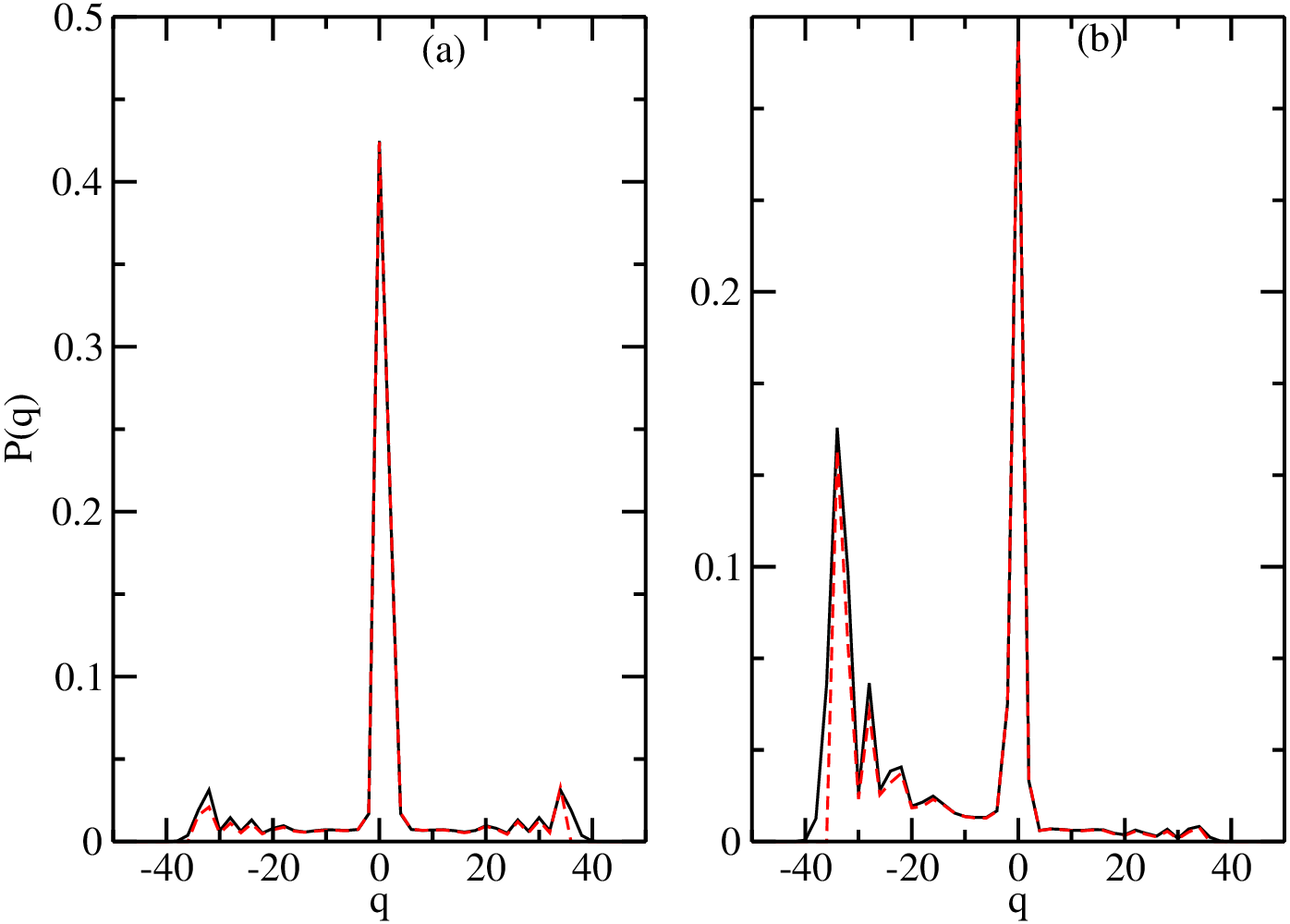}
\caption{(Color online) The probability distribution $P(q)$ for
detecting the walker at position $q$, for the initial coin state (a)
$\frac{1}{\sqrt{2}}(|h,y\rangle+|v,x\rangle)$ and
(b)$\frac{1}{\sqrt{2}}(|h,y\rangle-|v,x\rangle)$. The black solid
(red dashed) line is corresponding to exact (approximated) values.}
\label{fig3}
\end{figure}
The initial state of the coin can be selected as one of the states,
$|h,x\rangle$, $|h,y\rangle$, $|v,x\rangle$, $|v,y\rangle$, and
their possible superpositions. For a classical analog to this QRW we
look at the shift operator $S$ defined by Eq.(\ref{shift}). Clearly,
this walk is equivalent to the random walk of a single walker with
four dimensional coin where walker moves one step forward for two
possible outcomes of the coin and moves one step backward for the
other two. The probability of reaching the walker at position $q$
will be a binomial distribution
\begin{equation}
P(q)=\frac{n!}{\frac{n+q}{2}!\frac{n-q}{2}!}\frac{1}{2^n}.\label{binom}
\end{equation}
Thus classically we can not differentiate between the walk of single
photon in this case and the walk of a single walker with two sided
coin. In the following we show how QRW of single photon in this case
leads to different results than QRW with a two sided coin.

Here we present an approximate analysis of QRW for large number of
steps. We follow a similar method as discussed by Nayak et al
\cite{nayak} and Brun et al \cite{brun} using discrete spatial
Fourier transform and taking asymptotic limit of the Fourier
integrals for large number of steps. We relegate the details to the
appendix. For initial coin state $|h,x\rangle$, we get the final
state of the photon $|\Psi_{hx}(n)\rangle$ after $n$ steps as
follows.
\begin{eqnarray}
&&|\Psi_{hx}(n)\rangle=\sum_{q=-n}^{n}|\psi_{hx}(n,q)\rangle,\\
&&|\psi_{hx}(n,q)\rangle=f_{hx}|n,q,h,x\rangle+f_{hy}|n,q,h,y\rangle
+f_{vx}|n,q,v,x\rangle+f_{vy}|n,q,v,y\rangle, \label{psihx}
\end{eqnarray}
where the coefficients $f_{\mu\nu}$ are given by
\begin{eqnarray}
&&f_{hx}=\frac{1+(-1)^{n+q}}{8\pi}\left[\int_{-\pi}^{\pi}\frac{\cos(kq)}{2-\sqrt{2}\cos
k}dk+\sqrt{\frac{2\pi}{n|\phi''(k_0)|}}\frac{3C(n,q)-\sqrt{2}C(n-1,q+1)-\sqrt{2}C(n+1,q-1)}{1+\sin^2k_0}\right],\\
&&f_{hy}=\frac{1+(-1)^{n+q}}{8\pi}\left[\int_{-\pi}^{\pi}\frac{\sqrt{2}\cos[k(q+1)]-\cos(kq)}{2-\sqrt{2}\cos
k}dk+\sqrt{\frac{2\pi}{n|\phi''(k_0)|}}\frac{C(n,q)-\sqrt{2}C(n-1,q+1)}{1+\sin^2k_0}\right],\\
&&f_{vx}=\frac{1+(-1)^{n+q}}{8\pi}\left[\int_{-\pi}^{\pi}\frac{\cos[k(q+2)]}{2-\sqrt{2}\cos
k}dk-\sqrt{\frac{2\pi}{n|\phi''(k_0)|}}\frac{C(n,q+2)}{1+\sin^2k_0}\right],\\
&&f_{vy}=\frac{1+(-1)^{n+q}}{8\pi}\left[\int_{-\pi}^{\pi}\frac{\sqrt{2}\cos[k(q+1)]-\cos[k(q+2)]}{2-\sqrt{2}\cos
k}dk+\sqrt{\frac{2\pi}{n|\phi''(k_0)|}}\frac{\sqrt{2}C(n-1,q+1)-C(n,q)}{1+\sin^2k_0}\right],\\
&&C(n,q)=\cos(n\omega_0+qk_0+\pi/4),~~\omega_0=\omega_k|_{k=k_0},
\end{eqnarray}
where $\phi(k)=-(\omega_k+k\alpha)$, $\alpha=q/n$,
$\omega_k\in[\pi/4,3\pi/4]$ and defined as $\cos\omega_k=\cos
k/\sqrt{2}$, $k_0=\sin^{-1}(-\alpha/\sqrt{1-\alpha^2})$ and prime
denotes the derivative with respect to $k$. This approximation is
valid in the interval $\alpha\in[-1/\sqrt{2},1/\sqrt{2}]$, outside
of this interval $f_{\mu\nu}$ can be taken zero. In the expressions
of $f_{\mu\nu}$, the integral inside the bracket is independent of
$n$, and is responsible for the constant spikes in the probability
of detecting photon near initial position $q=0$. We evaluate this
integral numerically. The second term inside the bracket is sum of
cosines and completely characterizes QRW. Similarly, for initial
coin state $|h,y\rangle$, the state of the photon after $n$ steps is
\begin{eqnarray}
&&|\Psi_{hy}(n)\rangle=\sum_{q=-n}^{n}|\psi_{hy}(n,q)\rangle,\\
&&|\psi_{hy}(n,q)\rangle=g_{hx}|n,q,h,x\rangle+g_{hy}|n,q,h,y\rangle
+g_{vx}|n,q,v,x\rangle+g_{vy}|n,q,v,y\rangle, \label{psihy}
\end{eqnarray}
where
\begin{eqnarray}
&&g_{hx}=\frac{1+(-1)^{n+q}}{8\pi}\left[\int_{-\pi}^{\pi}\frac{\sqrt{2}\cos[k(q-1)]-\cos(kq)}{2-\sqrt{2}\cos
k}dk+\sqrt{\frac{2\pi}{n|\phi''(k_0)|}}\frac{C(n,q)-\sqrt{2}C(n+1,q-1)}{1+\sin^2k_0}\right],\\
&&g_{hy}=\frac{1+(-1)^{n+q}}{8\pi}\left[\int_{-\pi}^{\pi}\frac{(3-2\sqrt{2}\cos
k)\cos(kq)}{2-\sqrt{2}\cos k}dk+\sqrt{\frac{2\pi}{n|\phi''(k_0)|}}\frac{C(n,q)}{1+\sin^2k_0}\right],\\
&&g_{vx}=\frac{1+(-1)^{n+q}}{8\pi}\left[\int_{-\pi}^{\pi}\frac{\sqrt{2}\cos[k(q+1)]-\cos[k(q+2)]}{2-\sqrt{2}\cos
k}dk+\sqrt{\frac{2\pi}{n|\phi''(k_0)|}}\frac{\sqrt{2}C(n+1,q+1)-C(n,q)}{1+\sin^2k_0}\right],\\
&&g_{vy}=\frac{1+(-1)^{n+q}}{8\pi}\left[\int_{-\pi}^{\pi}\frac{2\cos
kq-2\sqrt{2}\cos[k(q+1)]+\cos[k(q+2)]}{2-\sqrt{2}\cos
k}dk-\sqrt{\frac{2\pi}{n|\phi''(k_0)|}}\frac{C(n,q)}{1+\sin^2k_0}\right].
\end{eqnarray}
From the symmetry of the arrangement (see Fig. \ref{fig1}), the
state of the photon after $n$ steps for initial coin states
$|v,x\rangle$ and $|v,y\rangle$ can be written by interchanging the
direction of propagation $h$ and $v$ and replacing $q$ by $-q$ in
$|\Psi_{hx}(n)\rangle$ and $|\Psi_{hy}(n)\rangle$ respectively.
\begin{eqnarray}
&&|\Psi_{vx}(n)\rangle=\sum_{q=-n}^{n}|\psi_{vx}(n,q)\rangle,\\
&&|\Psi_{vy}(n)\rangle=\sum_{q=-n}^{n}|\psi_{vy}(n,q)\rangle,\\
\label{psivx}
&&|\psi_{vx}(n,-q)\rangle=f_{hx}|n,-q,v,x\rangle+f_{hy}|n,-q,v,y\rangle
+f_{vx}|n,-q,h,x\rangle+f_{vy}|n,-q,h,y\rangle,\\
&&|\psi_{vy}(n,-q)\rangle=g_{hx}|n,-q,v,x\rangle+g_{hy}|n,-q,v,y\rangle
+g_{vx}|n,-q,h,x\rangle+g_{vy}|n,-q,h,y\rangle. \label{psivy}
\end{eqnarray}
For initial coin state as an arbitrary superposition
$\alpha|h,x\rangle+\beta|v,y\rangle$, the state of the photon after
$n$ steps is given by
\begin{equation}
|\Psi(n)\rangle=\alpha|\Psi_{hx}(n)\rangle+\beta|\Psi_{vy}(n)\rangle.
\end{equation}
In Figs. \ref{fig2} and \ref{fig3}, we have shown the probability
distribution for QRW of the single photon with initial coin states
$\frac{1}{\sqrt{2}}(|h,x\rangle\pm|v,y\rangle)$ and
$\frac{1}{\sqrt{2}}(|h,y\rangle\pm|v,x\rangle)$ after $50$ steps. We
plot the results using above approximate analysis as well as exact
numerical simulations. Clearly for large number of steps, say for
$n=50$, there are very small differences between the approximate
analysis and the exact simulations. Further for larger values of $n$
these differences will be negligible. It should be noted that the
probability distributions for detecting the photon at position $q$
in Figs. \ref{fig2} and \ref{fig3} are very much different than the
distributions for QRW with a two sided coin \cite{kempe}. In all
these cases the initial state is most probable state and the
distribution is sharply peaked at $q=0$. For initial coin states
$\frac{1}{\sqrt{2}}(|h,x\rangle+|v,y\rangle)$ and
$\frac{1}{\sqrt{2}}(|h,y\rangle+|v,x\rangle)$ the distributions are
symmetric but the side peaks are very small and most of the time
walker remains at its initial position very precisely. In the case
of QRW with initial coin states
$\frac{1}{\sqrt{2}}(|h,x\rangle-|v,y\rangle)$ and
$\frac{1}{\sqrt{2}}(|h,y\rangle-|v,x\rangle)$, in addition to the
narrow central peak, distributions have a peak along one side of the
position axis. Further for the state
$\frac{1}{\sqrt{2}}(|h,x\rangle-|v,y\rangle)$ the additional peak in
the distribution is along the positive side of the axis at
$q=n/\sqrt{2}$, while for state
$\frac{1}{\sqrt{2}}(|h,y\rangle-|v,x\rangle)$ the additional peak is
along the negative side at $q=-n/\sqrt{2}$. We emphasize that the
quantum random walk of a single photon depends very much on the
initial state, see for example the distinction between the Figs.
\ref{fig2}(a) and \ref{fig2}(b).
\section{QRW of Two photons with separable initial state}
In recent papers \cite{knight}, it has been shown that QRW is an
interference phenomenon and does not essentially depend on the
quantum nature of the state of the walker. As a result various
classical sources like low intensity lasers \cite{exp,bouwmeester}
and coherent state of radiation fields \cite{milburn,our} are used
to realize QRW.  In order to explore further the quantum nature of
random walk, we consider the case when two photons start QRW from a
separable initial state
\begin{equation}
|\Psi\rangle=|\Psi_1\rangle\otimes|\Psi_2\rangle,
\end{equation}
where $|\Psi\rangle$ is state of two photons and $|\Psi_1\rangle$
and $|\Psi_2\rangle$ are states of single photons. The state of the
photons after a certain number of steps is, in general, not a
separable state as quantum entanglement is produced by linear
optical elements. This is reminiscent of the well known property
\cite{mandel} of a 50-50 beam splitter where two incoming photons in
the separable state $|1,1\rangle$ go over to an entangled state of
the form $(|2,0\rangle+|0,2\rangle)/\sqrt{2}$. We first consider the
case of input states which are single photon states. We would also
consider the case when the input states are replaced by coherent
states
\subsection{QRW of two photons with initially in separable Fock states}
In our scheme, two photons act as two walkers. They enter in the
arrangement through the two input ports. Initially one photon
propagates in horizontal direction and the other in vertical
direction. We consider the initial state of the photons as one of
the four separable states $|0,0,h,x\rangle\otimes|0,0,v,x\rangle$,
$|0,0,h,y\rangle\otimes|0,0,v,y\rangle$,
$|0,0,h,x\rangle\otimes|0,0,v,y\rangle$, and
$|0,0,h,y\rangle\otimes|0,0,v,x\rangle$. We can write these states
in terms of initial field operators as follow.
\begin{eqnarray}
\label{sep1} |0,0,h,x\rangle\otimes|0,0,v,x\rangle\equiv
a^{(0,0)\dag}_{hx}a^{(0,0)\dag}_{vx}|0\rangle,\\
|0,0,h,x\rangle\otimes|0,0,v,y\rangle\equiv
a^{(0,0)\dag}_{hx}a^{(0,0)\dag}_{vy}|0\rangle,\\
|0,0,h,y\rangle\otimes|0,0,v,y\rangle\equiv
a^{(0,0)\dag}_{hy}a^{(0,0)\dag}_{vy}|0\rangle,\\
|0,0,h,y\rangle\otimes|0,0,v,x\rangle\equiv
a^{(0,0)\dag}_{hy}a^{(0,0)\dag}_{vx}|0\rangle, \label{sep4}
\end{eqnarray}
where $a^{(n,q)\dag}_{jk}$ is creation operator for a photon,
$a^{(n,q)\dag}_{jk}|0\rangle\equiv|n,q,j,k\rangle$, and $|0\rangle$
is vacuum. Here we present the analytical calculations for QRW of
few steps, say five steps. For linear optical elements, it is
sometimes instructive and transparent to work with the
transformation of operators. This is particularly so if the states
with more than one photons are involved. Thus for calculating final
state of the walkers after five steps, we express the initial field
operators, $a_{jk}^{(0,0)}$, in terms of the final field operators,
$a^{(5,q)}_{jk}$, after five steps.
\begin{eqnarray}
a_{hx}^{(0,0)}&=&\frac{1}{4\sqrt{2}}(a^{(5,-5)}_{vx}+a^{(5,-3)}_{hy}+a^{(5,-3)}_{vx}-a^{(5,-3)}_{vy}
+a^{(5,-1)}_{hx}\nonumber\\&+&3a^{(5,-1)}_{hy}+a^{(5,-1)}_{vx}+a^{(5,-1)}_{vy}+a^{(5,1)}_{hx}
+a^{(5,1)}_{hy}+a^{(5,1)}_{vx}\nonumber\\
&-&a^{(5,1)}_{vy}-3a^{(5,3)}_{hx}-a^{(5,3)}_{hy}+a^{(5,3)}_{vy}+a^{(5,5)}_{hx}),\label{trans1}\\
a_{hy}^{(0,0)}&=&\frac{1}{4\sqrt{2}}(-a^{(5,-5)}_{vx}-a^{(5,-3)}_{hy}+a^{(5,-3)}_{vx}+a^{(5,-3)}_{vy}
-a^{(5,-1)}_{hx}\nonumber\\&-&a^{(5,-1)}_{hy}+3a^{(5,-1)}_{vx}-3a^{(5,-1)}_{vy}+a^{(5,1)}_{hx}-a^{(5,1)}_{hy}
+a^{(5,1)}_{vx}\nonumber\\
&+&a^{(5,1)}_{vy}-a^{(5,3)}_{hx}-a^{(5,3)}_{hy}+a^{(5,3)}_{vy}+a^{(5,5)}_{hx}),\label{trans2}\\
a_{vx}^{(0,0)}&=&\frac{1}{4\sqrt{2}}(a^{(5,-5)}_{vx}+a^{(5,-3)}_{hy}-3a^{(5,-3)}_{vx}-a^{(5,-3)}_{vy}
+a^{(5,-1)}_{hx}\nonumber\\&-&a^{(5,-1)}_{hy}+a^{(5,-1)}_{vx}+a^{(5,-1)}_{vy}+a^{(5,1)}_{hx}
+a^{(5,1)}_{hy}+a^{(5,1)}_{vx}\nonumber\\
&+&3a^{(5,1)}_{vy}+a^{(5,3)}_{hx}-a^{(5,3)}_{hy}+a^{(5,3)}_{vy}+a^{(5,5)}_{hx}),\label{trans3}\\
a_{vy}^{(0,0)}&=&\frac{1}{4\sqrt{2}}(a^{(5,-5)}_{vx}+a^{(5,-3)}_{hy}-a^{(5,-3)}_{vx}-a^{(5,-3)}_{vy}
+a^{(5,-1)}_{hx}\nonumber\\&+&a^{(5,-1)}_{hy}+a^{(5,-1)}_{vx}-a^{(5,-1)}_{vy}+3a^{(5,1)}_{hx}
-3a^{(5,1)}_{hy}-a^{(5,1)}_{vx}\nonumber\\
&-&a^{(5,1)}_{vy}+a^{(5,3)}_{hx}+a^{(5,3)}_{hy}-a^{(5,3)}_{vy}-a^{(5,5)}_{hx}).\label{trans4}
\end{eqnarray}
Using transformations (\ref{trans1}) to (\ref{trans4}) and Eqs.
(\ref{sep1}) to (\ref{sep4}), the state of the quantum walkers,
corresponding to a particular initial state, after five steps can be
calculated. Clearly, the final state of the photons is an entangled
state and can not be expressed as a product of the states of two
single photons.
\begin{equation}
|\bar{\Psi}\rangle\neq|\bar{\Psi}_1\rangle\otimes|\bar{\Psi}_2\rangle,
\label{entangled}
\end{equation}
where $|\bar{\Psi}\rangle$ is final state of two photons and
$|\bar{\Psi}_1\rangle$ and $|\bar{\Psi}_2\rangle$ are final states
of single photons. It should be borne in mind that
Eqs.(\ref{trans1}) to (\ref{trans4}) should be supplemented by free
field operators at the open ports. These are important for the
operator algebra. However these do not contribute to the results
below and hence for brevity we have not written these explicitly in
Eqs.(\ref{trans1}) - (\ref{trans4}).

Finally, the results for calculated probability $P(q_1,q_2)$ for
detecting the walkers at positions $q_1$ and $q_2$ simultaneously,
after five steps, are shown in the Table.I. Note that the diagonal
elements $P(q,q)$ give the probability of finding two photons at the
site $q$. It is clear from the Table.I that the probability
distributions for all considered initial states are different to
each other. At the bottom of the table we present the probability
$P(q_1)$ of detecting {\it at least one} photon at the position
$q_1$, where $P(q_1)=\sum_{q_2}P(q_1,q_2)$. Further using the values
of $P(q_1,q_2)$ and $P(q_1)$, $P(q_2)$ for the positions $q_1$ and
$q_2$ we can also calculate the correlation
\begin{equation}
\sigma_{q_1q_2}=P(q_1,q_2)-P(q_1)P(q_2). \label{correlation}
\end{equation}
We found that the correlation $\sigma_{q_1q_2}$ is non zero almost
everywhere for all values of $P(q_1,q_2)$, which shows that though
initially walkers were in a separable state, but after few steps
their state is entangled. We emphasize that the correlation
(\ref{correlation}) is due to quantum as the state (\ref{entangled})
does not factorize. These correlations are due to the quantum nature
of the initial state and arise when the photons pass through the
linear optical elements. The origin of such correlations has been
observed by Mandel and coworkers \cite{mandel} in their pioneer work
on beam splitters. It should be noted here that if we replace the
photons with two coherent states the output state will be a
factorized state and the probability distribution will not exhibit
such correlations. Thus the QRW of two photons is completely
dependent of quantum nature of the state of the photons and no
coherent state can reproduce such QRW. It should be noted that the
normalization condition for $P(q_1,q_2)$ is given by $\sum_{q_1\geq
q_2}P(q_1,q_2)=1$. Further $\sum_{q_1}P(q_1)$ is not equal to $1$
and the normalization condition for $P(q_1)$ will be
$\sum_{q_1}P(q_1)+P(q_1,q_1)=2$. To see it more clearly consider the
following state of finding two photons at sites $1$ and $2$,
\begin{equation}
|\psi\rangle=\frac{1}{\sqrt{3}}\left(|1,1\rangle+|1,2\rangle+|2,2\rangle\right)
\label{examp}
\end{equation}
For this state the probabilities of detecting {\it at least one}
photon at site $1$ and $2$ are $P(1)=P(2)=2/3$ and the probabilities
of detecting both photons at same site are $P(1,1)=P(2,2)=1/3$ which
satisfy the above normalization condition.

\begin{table}
\caption{The calculated probabilities of detection for the photons
after five steps, for initial state (a)
$|0,0,h,x\rangle\otimes|0,0,v,x\rangle$, (b)
$|0,0,h,x\rangle\otimes|0,0,v,y\rangle$, (c)
$|0,0,h,y\rangle\otimes|0,0,v,y\rangle$, and (d)
$|0,0,h,y\rangle\otimes|0,0,v,x\rangle$. All values shown in the
table are $128$ times of the actual values.}

\begin{tabular}{c|c}
\begin{tabular}{|c||cccccc|}
\hline
\multicolumn{7}{|c|}{\rule[-3mm]{0mm}{8mm}{\bf(a)}}\\
$P(q_1,q_2)$ &$q_1=-5$ &$q_1=-3$ &$q_1=-1$ &$q_1=1$ &$q_1=3$ &$q_1=5$\\
\hline\hline
$q_2=-5$ & 0.25   & 1.5 &     2 &     2 &   1.5  &0.5\\
$q_2=-3$ & 1.5   & 4.25 &   18  &   10  &   16.5 &   1.5\\
$q_2=-1$ & 2   & 18&   6   &  20   &  10  &   2\\
$q_2=1$ & 2   & 10&   20  &    6  &   18 &    2\\
$q_2=3$ & 1.5   & 16.5&  10   &  18  &   4.25  &   1.5\\
$q_2=5$ & 0.5   & 1.5 &    2  &    2  &  1.5   & 0.25
\\\hline $P(q_1)$ &
7.75&51.75&58&58&51.75&7.75\\\hline
\end{tabular}
&
\begin{tabular}{|c|cccccc|}
\hline
\multicolumn{7}{|c|}{\rule[-3mm]{0mm}{8mm}{\bf(b)}}\\
$P(q_1,q_2)$ &$q_1=-5$ &$q_1=-3$ &$q_1=-1$ &$q_1=1$ &$q_1=3$ &$q_1=5$\\
\hline
$q_2=-5$ &      0.25   &   1   &   3   &   3   &   0.5   &   0\\
$q_2=-3$ &      1   &   1.25   &   7   &   9   &   4   &   0.5\\
$q_2=-1$ &      3   &   7   &   8   &  32   &   5   &   1\\
$q_2=1$  &      3   &   9   &  32   &  10   &  29   &   3\\
$q_2=3$  &      0.5   &   4   &   5   &  29   &   7.25   &   3\\
$q_2=5$  &      0   &   0.5   &   1   &   3   &   3   &   0.25\\
\hline $P(q_1)$ & 7.75&22.75&56&86&48.75&7.75\\\hline
\end{tabular}
\\\hline
\begin{tabular}{|c|cccccc|}
\hline
\multicolumn{7}{|c|}{\rule[-3mm]{0mm}{8mm}{\bf(c)}}\\
$P(q_1,q_2)$ &$q_1=-5$ &$q_1=-3$ &$q_1=-1$ &$q_1=1$ &$q_1=3$ &$q_1=5$\\
\hline
$q_2=-5$ &      0.25   &   1.5    &  2    &  2   &   1.5    &  0.5\\
$q_2=-3$ &      1.5   &   2.25    &  6    &  6   &   4.5    &  1.5\\
$q_2=-1$ &      2   &   6    & 12   & 56   &   6    &  2\\
$q_2=1$ &      2   &   6   & 56   & 12   &   6    &  2\\
$q_2=3$ &      1.5   &   4.5    &  6    &  6   &   2.25    &  1.5\\
$q_2=5$ &      0.5   &   1.5    &  2    &  2   &   1.5    &  0.25\\
\hline$P(q_1)$ & 7.75&21.75&84&84&21.75&7.75\\\hline
\end{tabular}
&
\begin{tabular}{|c|cccccc|}
\hline
\multicolumn{7}{|c|}{\rule[-3mm]{0mm}{8mm}{\bf(d)}}\\
$P(q_1,q_2)$ &$q_1=-5$ &$q_1=-3$ &$q_1=-1$ &$q_1=1$ &$q_1=3$ &$q_1=5$\\
\hline
$q_2=-5$ &      0.25   &   3   &   3   &   1   &   0.5   &   0\\
$q_2=-3$ &      3   &   7.25   &  29   &   5   &   4   &   0.5\\
$q_2=-1$ &      3   &  29   &  10   &  32   &   9   &   3\\
$q_2=1$ &      1   &   5   &  32   &   8   &   7   &   3\\
$q_2=3$ &      0.5   &   4   &   9  &   7   &   1.25   &   1\\
$q_2=5$ &      0   &   0.5   &   3   &   3   &   1   &   0.25\\
\hline $P(q_1)$ & 7.75&48.75&86&56&22.75&7.75\\\hline
\end{tabular}
\end{tabular}
\end{table}

\begin{figure*}
\begin{tabular}{c}
%\scalebox{1.3}{\includegraphics{FIG4.eps}}
\end{tabular}
\caption{(Color online) The probability $P(q_1,q_2)$ of detecting
the walkers at positions $q_1$ and $q_2$ after number of steps
$n=25$. The walkers come in the separable state (a)
$|0,0,h,x\rangle\otimes|0,0,v,x\rangle$, (b)
$|0,0,h,x\rangle\otimes|0,0,v,y\rangle$, (c)
$|0,0,h,y\rangle\otimes|0,0,v,y\rangle$, and (d)
$|0,0,h,y\rangle\otimes|0,0,v,x\rangle$.} \label{fig4}
\end{figure*}
After providing an approach to the analytical calculations for few
steps, we present numerical simulations for larger number of steps.
The operator $U=(S\otimes C)\otimes(S\otimes C)$ acting on the
initial state of the photons generates QRW of one step. Thus the
transformation $U^n$ will give the final state of the random walkers
after $n$ steps. Here unitary coin operator $C$ and the conditioned
shift operator $S$ for QRW are given by the Eqs. (\ref{coin}) and
(\ref{shift}) of the previous section.

In Fig. \ref{fig4} we have plotted the probability distributions for
detecting photons at positions $q_1$ and $q_2$ simultaneously after
$25$ steps of QRW. In Fig. \ref{fig4} the photons are in separable
initial states. We notice that each probability distribution is very
much different from the other. A very common feature in all plots is
sharp central peak which shows that in all cases the initial state
is the most probable state. In fact this is the property of
classical random walk, which has Gaussian probability distribution,
but in Fig. \ref{fig4} the central peak is much narrower than a
Gaussian distribution and the large spread of the distribution has
strickenly different behavior.

For approximate analytical calculations for large number of steps we
use the results (\ref{psihx}), (\ref{psihy}), (\ref{psivx}) and
(\ref{psivy}) in the following way. For example, for initial state
$|0,0,h,x\rangle\otimes|0,0,v,y\rangle$, the state of the photons
after $n$ steps will be
\begin{equation}
|\Psi_{xy}(n)\rangle=\frac{1}{\sqrt{2}}\sum_{q_1,q_2=-n}^{n}\left[|\psi_{hx}(n,q_1)\rangle\otimes
|\psi_{vy}(n,q_2)\rangle+|\psi_{hx}(n,q_2)\rangle\otimes|\psi_{vy}(n,q_1)\rangle\right].
\label{sep}
\end{equation}
Here $|\Psi_{xy}(n)\rangle$ has exchange degeneracy as both photons
are undistinguished after they have passed through the optical
arrangement. Similarly, we can calculate the final state of the
photons for other separable initial states shown in Fig.\ref{fig4}.
We have checked that the joint probability distributions
$P(q_1,q_2)$ for large number of steps calculated using above
approximate analysis  match with the exact simulations. Thus our
approximate analysis can be used to calculate the correlations
between the walkers after large number of steps very well.

Here we have explicitly shown that in our case QRW of two photons is
highly entangled, even though the photons start in separable states.
The entanglement is developed in the course of time when the photons
pass through the optical arrangement. In the next subsection we show
that the QRW of two photons in our case can not be produced by using
coherent states.
\subsection{QRW of two photons initially in separable coherent states}
Here we discuss the case of QRW when two photons in our scheme are
replaced with two weak coherent states. Let us consider that the
initial state is
\begin{equation}
|\psi(0)\rangle=|\alpha\rangle_{hx}^{(0,0)}|\beta\rangle_{vy}^{(0,0)},
\label{in}
\end{equation}
where indices to the coherent states $|\alpha\rangle$ and
$|\beta\rangle$ have their earlier assigned meanings. The initial
state in terms of field operators can be expressed as
\cite{coherent}
\begin{equation}
|\psi(0)\rangle=\exp(\alpha
a^{(0,0)\dag}_{hx}-\alpha^*a_{hx}^{(0,0)})\exp(\beta
a^{(0,0)\dag}_{vy}-\beta^*a_{vy}^{(0,0)})|0\rangle. \label{ex}
\end{equation}
Now using transformations (\ref{trans1}), (\ref{trans4}) and
Eq.(\ref{ex}) the final state, after five steps of QRW, is given by
\begin{eqnarray}
|\psi(5)\rangle=|\frac{\alpha+\beta}{4\sqrt{2}}\rangle^{(5,-5)}_{vx}|\frac{\alpha+\beta}{4\sqrt{2}}\rangle^{(5,-3)}_{hy}
|\frac{\alpha-\beta}{4\sqrt{2}}\rangle^{(5,-3)}_{vx}|\frac{-\alpha-\beta}{4\sqrt{2}}\rangle^{(5,-3)}_{vy}\nonumber\\
\otimes|\frac{\alpha+\beta}{4\sqrt{2}}\rangle^{(5,-1)}_{hx}|\frac{3\alpha+\beta}{4\sqrt{2}}\rangle^{(5,-1)}_{hy}
|\frac{\alpha+\beta}{4\sqrt{2}}\rangle^{(5,-1)}_{vx}|\frac{\alpha-\beta}{4\sqrt{2}}\rangle^{(5,-1)}_{vy}\nonumber\\
\otimes|\frac{\alpha+3\beta}{4\sqrt{2}}\rangle^{(5,1)}_{hx}|\frac{\alpha-3\beta}{4\sqrt{2}}\rangle^{(5,1)}_{hy}
|\frac{\alpha-\beta}{4\sqrt{2}}\rangle^{(5,1)}_{vx}|\frac{-\alpha-\beta}{4\sqrt{2}}\rangle^{(5,1)}_{vy}\nonumber\\
\otimes|\frac{-3\alpha+\beta}{4\sqrt{2}}\rangle^{(5,3)}_{hx}|\frac{-\alpha+\beta}{4\sqrt{2}}\rangle^{(5,3)}_{hy}
|\frac{\alpha-\beta}{4\sqrt{2}}\rangle^{(5,3)}_{vy}|\frac{\alpha-\beta}{4\sqrt{2}}\rangle^{(5,5)}_{hx},~~
\label{cstate}
\end{eqnarray}
where superscript $(5,q)$ on each state in (\ref{cstate}) denotes
the position of the photons at $q$ after $5$ steps. Clearly the
final state (\ref{cstate}) is a product state and the value of
correlation $\sigma_{q_1q_2}$ defined by (\ref{correlation}) will be
zero for such state.

Note that the probability of finding one photon in a coherent state
$|\alpha\rangle$ is $|\alpha|^2e^{-|\alpha|^2}$. For small enough
$|\alpha|$, it reduces to $|\alpha|^2$ which is the mean number of
photons in a coherent state. Thus, the normalized probability $P(q)$
of detecting a photon at $q$ in this case is equal to the average
number of photons detected at site $q$ divided by the average number
of incident photons $|\alpha|^2+|\beta|^2$.

In Table.II, we show the normalized probabilities of detecting
photon at site $q$ for small values of $\alpha$ and $\beta$. By
comparison of the cases corresponding to $\alpha=f$, $\beta=f$ and
$\alpha=f$, $\beta =0$, where $|f|\ll1$, we see how the interference
effects change these probabilities. Similarly for $\alpha=f$,
$\beta=f$ and $\alpha=0$, $\beta=f$, the probabilities depend on the
interference among various paths. Further these probabilities for
one photon detection are different from those listed in the Table.
I(b), for the case of two photons in separable Fock states
$|0,0,h,x\rangle\otimes|0,0,v,y\rangle$.
\begin{table}
\caption{The normalized probability $P(q)$ of detecting single
photon at particular position $q$, after QRW of $5$ steps with
initial state
$|\alpha\rangle_{hx}^{(0,0)}|\beta\rangle_{vy}^{(0,0)}$, for small
amplitudes $\alpha$ and $\beta$.}
\begin{tabular}{|c|c|cccccc|}
\hline
\multicolumn{8}{|c|}{\rule[-3mm]{0mm}{8mm}{Probability $P(q)$}}\\
$\alpha$ &$\beta$& $q=-5$ &$q=-3$ &$q=-1$ &$q=1$ &$q=3$ &$q=5$\\
\hline
$f$ & $f$  & 1/16 & 1/8  & 3/8 & 3/8 & 1/16 & 0  \\
$f$ & $-f$ & 0    & 1/16 & 1/8 & 3/8 & 3/8  & 1/16\\
$f$ & $0$  & 1/32 & 3/32 & 3/8 & 1/8 & 11/32& 1/32 \\
$0$ & $f$  & 1/32 & 3/32 & 1/8 & 5/8 & 3/32 & 1/32  \\\hline
\end{tabular}
\end{table}
It is noted that the normalized probability distribution of
detecting photons at particular position $q$ for such initial states
is identical to the case of single walker discussed in Sec.III, not
to the case of two walkers in Sec.IV A. For example, for $\alpha=f$
and $\beta=f$ the distribution is identical to Fig.\ref{fig2}(a) and
for $\alpha=f$, $\beta=-f$ we get distribution identical to
Fig.\ref{fig2}(b). This can be explained by examining the projection
of state (\ref{in}) in one photon space,
$|\psi_0\rangle\rightarrow\alpha|h,x\rangle\otimes|0\rangle+\beta|0\rangle\otimes|v,y\rangle$
as the contributions from spaces containing more than one photon are
of higher order in $\alpha$ and $\beta$.

\section{QRW of Two photons in entangled state}
Here we consider the case when photons start QRW in an entangled
state. Thus one photon enters from each input port and the state of
the photons is maximally entangled in polarization basis, as those
produced by a type-II downconverter. We consider that the state of
the photons at input ports is one of the four Bell's states
\cite{bell},
\begin{eqnarray}
|\psi^{\pm}\rangle=\frac{1}{\sqrt{2}}(|0,0,h,x\rangle\otimes|0,0,v,y\rangle
\pm|0,0,h,y\rangle\otimes|0,0,v,x\rangle),\label{bell1}\\
|\phi^{\pm}\rangle=\frac{1}{\sqrt{2}}(|0,0,h,x\rangle\otimes|0,0,v,x\rangle
\pm|0,0,h,y\rangle\otimes|0,0,v,y\rangle).\label{bell2}
\end{eqnarray}
The states (\ref{bell1}) and (\ref{bell2}) can be written, in terms
of the field operators, as
\begin{eqnarray}
|\psi^{\pm}\rangle=\frac{1}{\sqrt{2}}(a^{(0,0)\dag}_{hx}a^{(0,0)\dag}_{vy}\pm
a^{(0,0)\dag}_{hy}a^{(0,0)\dag}_{vx})|0\rangle,\label{input1}\\
|\phi^{\pm}\rangle=\frac{1}{\sqrt{2}}(a^{(0,0)\dag}_{hx}a^{(0,0)\dag}_{vx}\pm
a^{(0,0)\dag}_{hy}a^{(0,0)\dag}_{vy})|0\rangle,\label{input2}
\end{eqnarray}
Following a similar procedure as discussed in Sec.IV, we can
calculate the final state of photons after five steps using
transformations (\ref{trans1}) to (\ref{trans4}) and Eqs.
(\ref{input1}) to (\ref{input2}). In this case also we find that the
final state of the photons is an entangled state. Thus in this case
both initial and final states of the walkers are entangled state and
no classical analog for such states is possible.

In Table. III, we show the results for detecting the photons
simultaneously when they start QRW from one of the Bell's states
(\ref{input1}) and (\ref{input2}). We notice that the differences
between the distributions for symmetric states and antisymmetric
states are much larger than the differences between the
distributions for symmetric-symmetric or antisymmetric-antisymmetric
states. In terms of correlation (\ref{correlation}), both the
initial and the final states are highly correlated.
\begin{table}
\caption{The calculated probabilities of detection for the photons
after five steps, for initial state (a)
$1/\sqrt{2}\left(|0,0,h,x\rangle\otimes|0,0,v,y\rangle+|0,0,h,y\rangle\otimes|0,0,v,x\rangle\right)$,
(b)
$1/\sqrt{2}\left(|0,0,h,x\rangle\otimes|0,0,v,y\rangle-|0,0,h,y\rangle\otimes|0,0,v,x\rangle\right)$,
(c)
$1/\sqrt{2}\left(|0,0,h,x\rangle\otimes|0,0,v,x\rangle+|0,0,h,y\rangle\otimes|0,0,v,y\rangle\right)$,
and (d)
$1/\sqrt{2}\left(|0,0,h,x\rangle\otimes|0,0,v,x\rangle-|0,0,h,y\rangle\otimes|0,0,v,y\rangle\right)$.
All values shown in the table are $128$ times of the actual values.}

\begin{tabular}{c|c}
\begin{tabular}{|c|cccccc|}
\hline
\multicolumn{7}{|c|}{\rule[-3mm]{0mm}{8mm}{\bf(a)}}\\
$P(q_1,q_2)$ &$q_1=-5$ &$q_1=-3$ &$q_1=-1$ &$q_1=1$ &$q_1=3$ &$q_1=5$\\
\hline
$q_2=-5$ &  0 & 1 & 3  & 3 & 1 & 0  \\
$q_2=-3$ &  1 & 4 & 15 & 7 & 8 & 1  \\
$q_2=-1$ &  3 & 15& 9  & 34& 7 & 3  \\
$q_2=1$  &  3 & 7 & 34 & 9 & 15& 3  \\
$q_2=3$  &  1 & 8 & 7  & 15& 4 & 1  \\
$q_2=5$  &  0 & 1 & 3  & 3 & 1 & 0  \\
\hline $P(q_1)$ & 8 & 36 &71 & 71 & 36&8\\\hline
\end{tabular}&

\begin{tabular}{|c|cccccc|}
\hline
\multicolumn{7}{|c|}{\rule[-3mm]{0mm}{8mm}{\bf(b)}}\\
$P(q_1,q_2)$ &$q_1=-5$ &$q_1=-3$ &$q_1=-1$ &$q_1=1$ &$q_1=3$ &$q_1=5$\\
\hline
$q_2=-5$ & 0.5& 3 & 3  & 1 & 0 & 0  \\
$q_2=-3$ &  3 &4.5& 21 & 7 & 0 & 0  \\
$q_2=-1$ &  3 & 21& 9  & 30& 7 & 1  \\
$q_2=1$  &  1 & 7 & 30 & 9 & 21& 3  \\
$q_2=3$  &  0 & 0 & 7  & 21&4.5& 3  \\
$q_2=5$  &  0 & 0 & 1  & 3 & 3 & 0.5  \\
\hline $P(q_1)$ & 7.5&35.5&71&71&35.5&7.5\\\hline
\end{tabular}\\
\hline\hline
\begin{tabular}{|c|cccccc|}
\hline
\multicolumn{7}{|c|}{\rule[-3mm]{0mm}{8mm}{\bf(c)}}\\
$P(q_1,q_2)$ &$q_1=-5$ &$q_1=-3$ &$q_1=-1$ &$q_1=1$ &$q_1=3$ &$q_1=5$\\
\hline
$q_2=-5$ &  0 & 0 & 1  & 3 & 3 & 1  \\
$q_2=-3$ &  0 & 2 & 7  & 9 &17 & 3  \\
$q_2=-1$ &  1 & 7 & 9  & 42& 9 & 3  \\
$q_2=1$  &  3 & 9 & 42 & 9 & 7 & 1  \\
$q_2=3$  &  3 &17 & 9  & 7 & 2 & 0  \\
$q_2=5$  &  1 & 3 & 3  & 1 & 0 & 0  \\\hline $P(q_1)$ &
8&38&71&71&38&8\\\hline
\end{tabular}&

\begin{tabular}{|c|cccccc|}
\hline
\multicolumn{7}{|c|}{\rule[-3mm]{0mm}{8mm}{\bf(d)}}\\
$P(q_1,q_2)$ &$q_1=-5$ &$q_1=-3$ &$q_1=-1$ &$q_1=1$ &$q_1=3$ &$q_1=5$\\
\hline
$q_2=-5$ & 0.5& 3 & 3  & 1 & 0 & 0  \\
$q_2=-3$ &  3 &4.5& 17 & 7 & 4 & 0  \\
$q_2=-1$ &  3 & 17& 9  & 34& 7 & 1  \\
$q_2=1$  &  1 & 7 & 34 & 9 &17 & 3  \\
$q_2=3$  &  0 & 4 & 7  & 17&4.5& 3  \\
$q_2=5$  &  0 & 0 & 1  & 3 & 3 & 0.5  \\\hline $P(q_1)$ &
7.5&35.5&71&71&35.5&7.5\\\hline
\end{tabular}
\end{tabular}
\end{table}

\begin{figure*}
\begin{tabular}{c}
%\scalebox{1.3}{\includegraphics{FIG5.eps}}
\end{tabular}
\caption{(Color online) The probability $P(q_1,q_2)$ of detecting
the walkers at positions $q_1$ and $q_2$ after number of steps
$n=25$. The walkers come in the entangled state (a)
$|\psi^{+}\rangle$, (b) $|\psi^{-}\rangle$, (c) $|\phi^{+}\rangle$,
and (d) $|\phi^{-}\rangle$.} \label{fig5}
\end{figure*}

\begin{figure}
\centering
\includegraphics[width=3in]{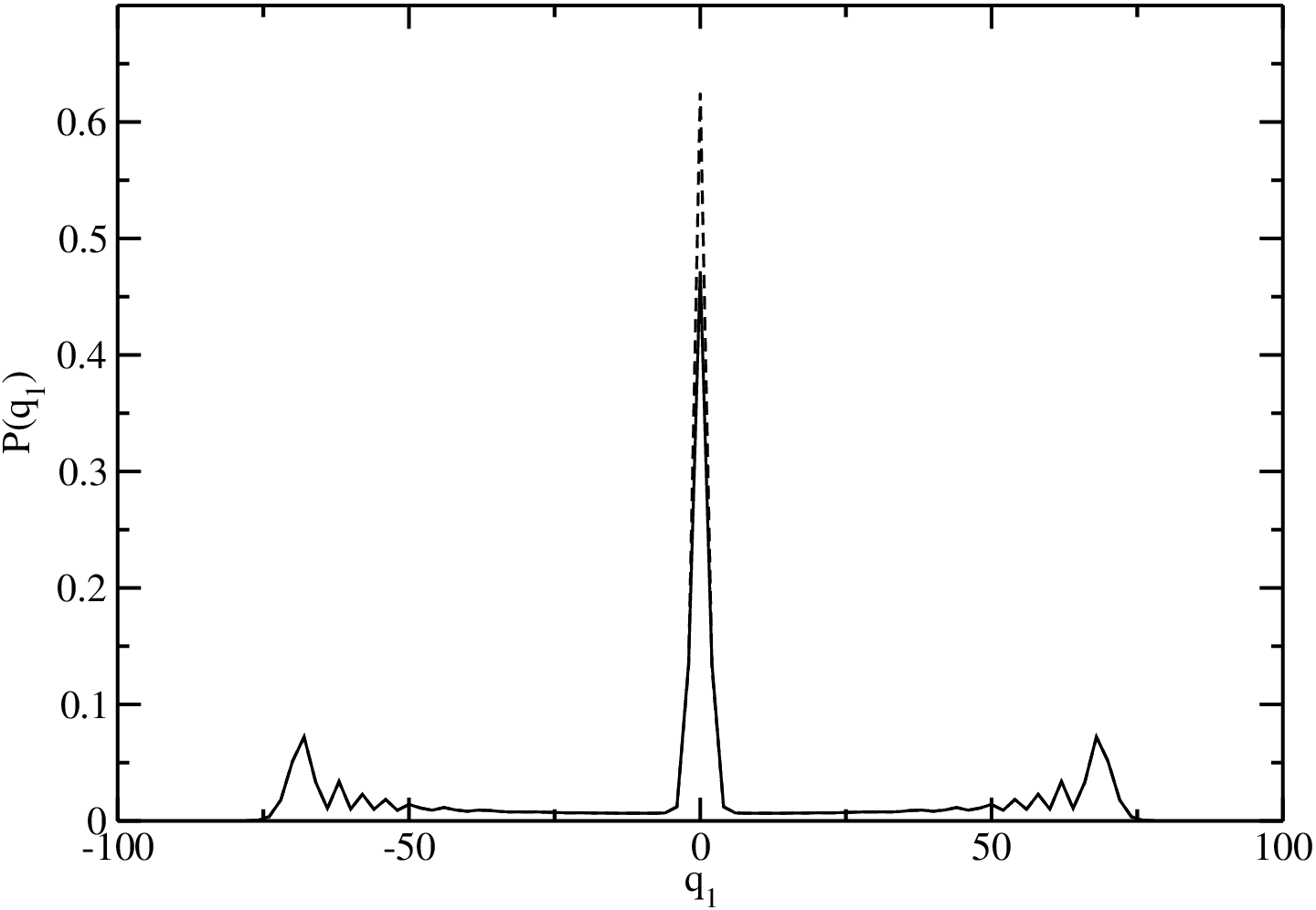}
\caption{The probability distribution $P(q_1)$ for detecting {\it at
least one} walker at position $q_1$ after $100$ steps. For initial
states $|\psi^{\pm}\rangle$ (dotted line) and $|\phi^{\pm}\rangle$
(solid line), the distribution remains almost same.} \label{fig6}
\end{figure}
For larger number of steps, we do numerical simulations following
the method discussed in Sec.IV A. In Fig. \ref{fig5} the initial
states of the photons are maximally entangled states (\ref{bell1})
and (\ref{bell2}). For different entangled states we get different
probability distributions. The differences between the distributions
for symmetric states and antisymmetric states are much larger than
the differences between the distributions for symmetric-symmetric or
antisymmetric-antisymmetric states. Further, we have seen by direct
computation that the probability distributions for entangled states
can not be reproduced exactly by the walkers having initial states
as an incoherent mixture of separable states. In Fig. \ref{fig6} we
have shown the probability distribution $P(q_1)$ for detecting {\it
at least one} walker at the position $q_1$. In this case the
distribution is symmetric for both negative and positive values of
$q_1$ and has a sharp maxima at initial position $q_1=0$. The
probability distribution for walkers with maximally entangled
initial states (\ref{bell1}) and (\ref{bell2}) remains almost
invariant except at initial position. At initial position the
probability of detecting one walker is larger for states
(\ref{bell1}) than the case of initial state (\ref{bell2}). In the
case of two-photon QRW with separable Fock states discussed in Sec.
IV A, the probability $P(q_1)$ is different for different initial
state. For initial states $|0,0,h,x\rangle\otimes|0,0,v,x\rangle$
and $|0,0,h,y\rangle\otimes|0,0,v,y\rangle$ the distribution
$P(q_1)$ is symmetric on both sides of $q_1$-axis and side peaks
appears in both positive and negative directions, while in the case
of $|0,0,h,x\rangle\otimes|0,0,v,y\rangle$ and
$|0,0,h,y\rangle\otimes|0,0,v,x\rangle$ the probability
distributions are antisymmetric with one side peak in positive and
negative direction respectively.

For large number of steps, we can again use our approximate results
(\ref{psihx}), (\ref{psihy}), (\ref{psivx}) and (\ref{psivy}) in the
following way. For initial state as one of the Bell's state
$|\psi^{\pm}\rangle$ the final state of the photons after $n$ steps
will be
\begin{eqnarray}
|\Psi^{\pm}(n)\rangle=\frac{1}{2}\sum_{q_1,q_2=-n}^{n}\left[|\psi_{hx}(n,q_1)\rangle\otimes
|\psi_{vy}(n,q_2)\rangle+|\psi_{hx}(n,q_2)\rangle\otimes|\psi_{vy}(n,q_1)\rangle\right.\nonumber\\
\left.\pm|\psi_{hy}(n,q_1)\rangle\otimes
|\psi_{vx}(n,q_2)\rangle\pm|\psi_{hy}(n,q_2)\rangle\otimes|\psi_{vx}(n,q_1)\rangle\right].
\label{bel1}
\end{eqnarray}
Similarly, for initial state as Bell's states $|\phi^{\pm}\rangle$,
the state of the photons will be
\begin{eqnarray}
|\Phi^{\pm}(n)\rangle=\frac{1}{2}\sum_{q_1,q_2=-n}^{n}\left[|\psi_{hx}(n,q_1)\rangle\otimes
|\psi_{vx}(n,q_2)\rangle+|\psi_{hx}(n,q_2)\rangle\otimes|\psi_{vx}(n,q_1)\rangle\right.\nonumber\\
\left.\pm|\psi_{hy}(n,q_1)\rangle\otimes
|\psi_{vy}(n,q_2)\rangle\pm|\psi_{hy}(n,q_2)\rangle\otimes|\psi_{vy}(n,q_1)\rangle\right].
\label{bel2}
\end{eqnarray}
The results (\ref{bel1}) and (\ref{bel2}) match very well with exact
simulations for large number of steps. Here it should be noted that
QRW in our case is different than the case of two entangled walkers
discussed in Ref. \cite{bose2}. In Ref. \cite{bose2}, the
correlation between the walkers is completely because of their
correlated initial state only. The walkers are nonidentical and
evolve completely independent to each other. Thus if the initial
state will be a separable state there will be no correlation in QRW
and can be produced using walkers in coherent states. It should be
borne in mind that we can not construct entangled state of one
x-polarized photon and the other y-polarized photon unless some
other degree of freedom like direction of propagation is introduced.
\section{Conclusions}
In conclusion, we have discussed how the quantum features of QRW can
be uncovered by studying QRW by photons in a number of quantum
states and by detecting the coincidence correlations between two
photons. We use two photons, each of which can exist in two
polarization states. Further, the photons can travel in either
vertical or horizontal direction. The walk is realized by optical
elements consisting of polarization beam splitters and half wave
plates. We first consider the QRW by a single photon and show how
the result of QRW depends on initial state of the photon. A
comparison of the Figs.\ref{fig2}(a) and \ref{fig2}(b) shows that
the probabilities depend on the relative phases in the initial
superposition state of the photon. We next consider QRW by two
walkers in separable quantum states like Fock states. We show that
our optical arrangement entangles two photons even though initially
they are in separable states. The joint probability of finding the
two photons at two different sites now depends on the initial
separable quantum state. Further the probability for single photon
detection is different from those calculated in Sec.III. We also
examine the QRW of two photons in coherent states. In this case the
state of the two photons remains separable. The single photon
detection probabilities depend on the initial amplitudes of the two
coherent states and thus the interferences are prominent. Finally we
consider QRW of two photons in all four Bell states. The resulting
joint probability distributions are quite different from those
calculated for separable states. All the cases discussed reveal very
interesting quantum character. Clearly our analysis is applicable to
other systems as well, for example electrons with appropriate
arrangement of Stern-Gerlach fields.

\acknowledgments
We thank NSF grant no CCF-0524673 for supporting
this work.

\appendix
\section{Fourier Analysis of QRW of a Single Photon}
We can write Eq.(\ref{evol}) for the wave function of the single
photon in the form
\begin{equation}
\psi(n+1,q)=M_+\psi(n,q-1) +M_-\psi(n,q+1) \label{eq1}
\end{equation}
where
\begin{eqnarray}
M_+=\frac{1}{\sqrt{2}}\left[\begin{array}{cccc}
1&1&0&0\\0&0&1&-1\\0&0&0&0\\0&0&0&0\end{array}\right],\\
M_-=\frac{1}{\sqrt{2}}\left[\begin{array}{cccc}
0&0&0&0\\0&0&0&0\\0&0&1&1\\1&-1&0&0\end{array}\right],
\end{eqnarray}
and the state of the photon after $n$ steps,
$|\psi(n,q)\rangle=c_{hx}|n,q,h,x\rangle+c_{hy}|n,q,h,y\rangle
+c_{vx}|n,q,v,x\rangle+c_{vy}|n,q,v,y\rangle$, is expressed in the
matrix form
\begin{equation}
\psi(n,q)=\left[\begin{array}{c}c_{hx}(n,q)\\c_{hy}(n,q)\\c_{vx}(n,q)\\c_{vy}(n,q)
\end{array}\right].
\end{equation}
Now we solve Eq.(\ref{eq1}) using spatial discrete Fourier
transform. The spatial discrete Fourier transform for
$k\in[-\pi,\pi]$ is defined by
\begin{eqnarray}
\tilde{\psi}(k)=\sum_x\psi(x)e^{ikx}\label{df},\\
{\rm
and}~\psi(x)=\frac{1}{2\pi}\int_{-\pi}^{\pi}\tilde{\psi}(k)e^{-ikx}dk.
\label{idf}
\end{eqnarray}
We write Eq.(\ref{eq1})in the Fourier domain using transform
(\ref{df}) as following,
\begin{eqnarray}
\tilde{\psi}(n+1,k)&=&\sum_xM_+\psi(n,x-1)e^{ikx}
+M_-\psi(n,x+1)e^{ikx}\nonumber\\
&=&(e^{ik}M_++e^{-ik}M_-)\tilde{\psi}(n,k)\nonumber\\
&=&M_k\tilde{\psi}(n,k) \label{eq2}
\end{eqnarray}
where \begin{equation}
M_k=\frac{1}{\sqrt{2}}\left[\begin{array}{cccc}
e^{ik}&e^{ik}&0&0\\0&0&e^{ik}&-e^{ik}\\0&0&e^{-ik}&e^{-ik}\\e^{-ik}&-e^{-ik}&0&0\end{array}\right].
\end{equation}
It is clear from Eq.(\ref{eq2}) that in the Fourier domain the $n$
step QRW of the photon starting from the state $\tilde{\psi}(0,k)$
is given by
\begin{eqnarray}
\tilde{\psi}(n,k)&=&(M_k)^n\tilde{\psi}(0,k)\nonumber\\
&=&\sum_m(\lambda_m)^n|\Phi_m\rangle\langle\Phi_m|\tilde{\psi}(0,k)\rangle
\label{fwalk}
\end{eqnarray}
where $\lambda_m$ and $\Phi_m$ are eigenvalue and corresponding
eigenvector of $M_k$. Thus, we can calculate exact value of
$\tilde{\psi}(n,k)$ by diagonalizing $M_k$ and writing initial state
of the photon in Fourier domain. The state of the photon after $n$
steps in original domain $\psi(n,q)$ is given by the inverse Fourier
transform of $\tilde{\psi}(n,k)$, defined by Eq.(\ref{idf}). The
eigenvalues of $M_k$ are $\lambda_1=-1$, $\lambda_2=1$,
$\lambda_3=e^{-i\omega_k}$ and $\lambda_4=e^{i\omega_k}$, where
$\omega_k\in[\pi/4,3\pi/4]$ and defined as $\cos\omega_k=\cos
k/\sqrt{2}$. The corresponding eigenvectors are
\begin{eqnarray}
&&|\Phi_1\rangle=\frac{1}{2}\sqrt{\frac{3+2\sqrt{2}\cos
k}{2+\sqrt{2}\cos k}} \left[\begin{array}{c}
\frac{-e^{ik}}{\sqrt{2}+e^{-ik}}\\
\frac{\sqrt{2}+e^{ik}}{\sqrt{2}+e^{-ik}}\\
\frac{-e^{-ik}}{\sqrt{2}+e^{-ik}}\\1\end{array}\right],\nonumber\\
&&|\Phi_2\rangle=\frac{1}{2}\sqrt{\frac{3-2\sqrt{2}\cos
k}{2-\sqrt{2}\cos k}} \left[\begin{array}{c}
\frac{e^{ik}}{\sqrt{2}-e^{-ik}}\\
\frac{\sqrt{2}-e^{ik}}{\sqrt{2}-e^{-ik}}\\
\frac{e^{-ik}}{\sqrt{2}-e^{-ik}}\\1\end{array}\right],\nonumber\\
&&|\Phi_3\rangle=\frac{1}{2\sqrt{1+\sin^2k}}\left[\begin{array}{c}-1+\sqrt{2}e^{-i(\omega_k-k)}\\
-1\\1-\sqrt{2}e^{-i(\omega_k+k)}\\1\end{array}\right],\nonumber\\
&&|\Phi_4\rangle=\frac{1}{2\sqrt{1+\sin^2k}}\left[\begin{array}{c}-1+\sqrt{2}e^{i(\omega_k+k)}\\
-1\\1-\sqrt{2}e^{i(\omega_k-k)}\\1\end{array}\right].\nonumber
\end{eqnarray}
If the initial state of the photon is $|0,0,h,x\rangle$ the
corresponding state in fourier domain will be
$\tilde{\psi}(0,k)=[1,0,0,0]^{T}$ for all values of $k$. Similarly
if the photon starts QRW from state $|0,0,h,y\rangle$ the initial
state in Fourier domain will be $[0,1,0,0]^T$. Now from
Eq.(\ref{fwalk}), we calculate the state of the quantum walker
$\tilde{\psi}(n,k)$ in fourier domain. Taking the inverse Fourier
transform of $\tilde{\psi}(n,k)$, we get the state of the walker
after $n$ steps $\psi(n,q)$ in the original coordinate space. For
initial state of the photon $|0,0,h,x\rangle$, we get
\begin{equation}
|\psi_{hx}(n,q)\rangle=f_{hx}|n,q,h,x\rangle+f_{hy}|n,q,h,y\rangle
+f_{vx}|n,q,v,x\rangle+f_{vy}|n,q,v,y\rangle,
\end{equation}
where the coefficients $f_{\alpha\beta}$ are given by
\begin{eqnarray}
&&f_{hx}=\frac{1+(-1)^{n+q}}{8\pi}\left[\int_{-\pi}^{\pi}\frac{\cos(kq)}{2-\sqrt{2}\cos
k}dk+\int_{-\pi}^{\pi}\frac{3-2\sqrt{2}\cos(\omega_k-k)}{1+\sin^2k}e^{-i(n\omega_k+kq)}dk\right],\\
&&f_{hy}=\frac{1+(-1)^{n+q}}{8\pi}\left[\int_{-\pi}^{\pi}\frac{\sqrt{2}\cos[k(q+1)]-\cos(kq)}{2-\sqrt{2}\cos
k}dk+\int_{-\pi}^{\pi}\frac{1-\sqrt{2}e^{i(\omega_k-k)}}{1+\sin^2k}e^{-i(n\omega_k+kq)}dk\right],\\
&&f_{vx}=\frac{1+(-1)^{n+q}}{8\pi}\left[\int_{-\pi}^{\pi}\frac{\cos[k(q+2)]}{2-\sqrt{2}\cos
k}dk-\int_{-\pi}^{\pi}\frac{e^{-2ik}}{1+\sin^2k}e^{-i(n\omega_k+kq)}dk\right],\\
&&f_{vy}=\frac{1+(-1)^{n+q}}{8\pi}\left[\int_{-\pi}^{\pi}\frac{\sqrt{2}\cos[k(q+1)]-\cos[k(q+2)]}{2-\sqrt{2}\cos
k}dk+\int_{-\pi}^{\pi}\frac{\sqrt{2}e^{i(\omega_k-k)}-1}{1+\sin^2k}e^{-i(n\omega_k+kq)}dk\right].
\end{eqnarray}
Similarly, for initial state of the photon $|0,0,h,y\rangle$, the
state of the walker after $n$ steps is
\begin{equation}
|\psi_{hy}(n,q)\rangle=g_{hx}|n,q,h,x\rangle+g_{hy}|n,q,h,y\rangle
+g_{vx}|n,q,v,x\rangle+g_{vy}|n,q,v,y\rangle,
\end{equation}
where
\begin{eqnarray}
&&g_{hx}=\frac{1+(-1)^{n+q}}{8\pi}\left[\int_{-\pi}^{\pi}\frac{\sqrt{2}\cos[k(q-1)]-\cos(kq)}{2-\sqrt{2}\cos
k}dk+\int_{-\pi}^{\pi}\frac{1-\sqrt{2}e^{-i(\omega_k-k)}}{1+\sin^2k}e^{-i(n\omega_k+kq)}dk\right],\\
&&g_{hy}=\frac{1+(-1)^{n+q}}{8\pi}\left[\int_{-\pi}^{\pi}\frac{(3-2\sqrt{2}\cos
k)\cos(kq)}{2-\sqrt{2}\cos k}dk+\int_{-\pi}^{\pi}\frac{e^{-i(n\omega_k+kq)}}{1+\sin^2k}dk\right],\\
&&g_{vx}=\frac{1+(-1)^{n+q}}{8\pi}\left[\int_{-\pi}^{\pi}\frac{\sqrt{2}\cos[k(q+1)]-\cos[k(q+2)]}{2-\sqrt{2}\cos
k}dk+\int_{-\pi}^{\pi}\frac{\sqrt{2}e^{-i(\omega_k+k)}-1}{1+\sin^2k}e^{-i(n\omega_k+kq)}dk\right],\\
&&g_{vy}=\frac{1+(-1)^{n+q}}{8\pi}\left[\int_{-\pi}^{\pi}\frac{2\cos
kq-2\sqrt{2}\cos[k(q+1)]+\cos[k(q+2)]}{2-\sqrt{2}\cos
k}dk-\int_{-\pi}^{\pi}\frac{e^{-i(n\omega_k+kq)}}{1+\sin^2k}dk\right].
\end{eqnarray}
Because of the factor $1+(-1)^{n+q}$, coefficients $f_{\alpha\beta}$
and $g_{\alpha\beta}$ are nonzero for even values of $n+q$ only,
which is corresponding to the positions of detecting photon in the
arrangement (see Fig.\ref{fig1}) for a fixed value of $n$. The
integrals inside the bracket can not be evaluated exactly. Further
the first integral inside the bracket is independent of $n$, and is
responsible for the constant spikes in the probability of detecting
photon near the initial position $q=0$. The second integral inside
the bracket completely characterizes QRW.

In order to understand the nature of QRW we do asymptotic analysis
of the second integral inside the bracket in the coefficients
$f_{\alpha\beta}$ and $g_{\alpha\beta}$ as follows. The integral has
the form of
\begin{equation}
I(n)=\int_{-\pi}^{\pi}f(k)e^{in\phi(k)}dk,
\end{equation}
where $\phi(k)=-(\omega_k+k\alpha)$, $\alpha=q/n$, and $f(k)$ is
function of $k$ only. In the limit of large $n$ we use stationary
phase approximation \cite{bornwolf} and find the approximate value
of $I(n)$.
\begin{equation}
I(n)\approx
\sqrt{\frac{2\pi}{n|\phi''(k_0)|}}\Re\left\{f(k_0)e^{in\phi(k_0)-i\pi/4}\right\},~{\rm
for~}-\frac{1}{\sqrt{2}}<\alpha<\frac{1}{\sqrt{2}},
\end{equation}
where $k_0=\sin^{-1}(-\alpha/\sqrt{1-\alpha^2})$ and prime denotes
the derivative with respect to $k$. For $\alpha=\pm1/\sqrt{2}$,
$\phi(k)$ has stationary point of order two at $k=\pm\pi/2$, i.e.
$\phi'(\pm\pi/2)=0,~\phi''(\pm\pi/2)=0,~\phi'''(\pm\pi/2)\neq0$. At
these points
\begin{eqnarray}
I(n)\approx
\left(\frac{6}{n}\right)^{1/3}\frac{\sqrt{2}\Gamma(1/3)}{\sqrt{3}}f(-\pi/2)e^{in\phi(-\pi/2)},~{\rm
for~}\alpha=\frac{1}{\sqrt{2}},\\
I(n)\approx
\left(\frac{6}{n}\right)^{1/3}\frac{\sqrt{2}\Gamma(1/3)}{\sqrt{3}}f(\pi/2)e^{in\phi(\pi/2)},~{\rm
for~}\alpha=-\frac{1}{\sqrt{2}}.
\end{eqnarray}
Everywhere else $I(n)$ has no stationary point and averages to zero
for $n\rightarrow\infty$.


\begin{thebibliography}{99}
\bibitem{aharonov}Y. Aharonov, L. Davidovich, and N. Zagury,
Phys. Rev. A {\bf48}, 1687 (1993).
\bibitem{sanders}B. C. Sanders, S. D. Bartlett, B. Tregenna, and P. L.
Knight, Phys. Rev. A {\bf67}, 042305 (2003); V. Kendon and B. C.
Sanders, Phys. Rev. A {\bf71}, 022307 (2005) .
\bibitem{zubairy}T. Di, M. Hillery, and M. S. Zubairy,
Phys. Rev. A {\bf70}, 032304 (2004); M. Hillery, J. Bergou, and E.
Feldman, Phys. Rev. A {\bf68}, 032314 (2003).
\bibitem{milburn}B. C. Travaglione and G. J. Milburn,
Phys. Rev. A {\bf65}, 032310 (2002).
\bibitem{nmr}C. A. Ryan, M. Laforest, J. C. Boileau, and R.
Laflamme, Phys. Rev. A {\bf72}, 062317 (2005).
\bibitem{chaos}P. Ribeiro, P. Milman, and R. Mosseri, Phys. Rev. Lett. {\bf93}, 190503
(2004).
\bibitem{kempe}J. Kempe, Contemp. Phys. {\bf44}, 307 (2003).
\bibitem{exp}B. Do, M. L. Stohler, S. Balasubramanian, D. S.
Elliott, C. Eash, E. Fischbach, M. A. Fischbach, A. Mills, B.
Zwickl, J. Opt. Soc. Am. B, {\bf22}, 499 (2005).
\bibitem{zhao}Z. Zhao, J. Du, H. Li, T. Yang, Z.-B. Chen, and J.-W. Pan, e-print quant-ph/0212149.
\bibitem{jeong} H. Jeong, M. Paternostro, and M. S. Kim, Phys. Rev. A {\bf69} , 012310
(2004).
\bibitem{knight}P. L. Knight, E. Roldan, and J. E. Sipe,
Phys. Rev. A {\bf68}, 020301(R) (2003); Opt. Commun. {\bf227}, 147
(2003).
\bibitem{ent-knight}I. Carneiro, M. Loo, X. Xu, M. Girerd, V. Kendon, and P. L.
Knight, New J. Phys. {\bf7}, 156 (2005).
\bibitem{bouwmeester}D. Bouwmeester, I. Marzoli, G. P. Karman, W. Schleich, and J. P.
Woerdman, Phys. Rev. A {\bf61}, 013410 (2000).
\bibitem{our}G. S. Agarwal and P. K. Pathak, Phys. Rev. A {\bf72}, 033815
(2005).
\bibitem{bose1}S. E. Venegas-Andraca, J. L. Ball, K. Burnett, and S. Bose, New J. Phys.
{\bf7}, 221 (2005).
\bibitem{bose2}Y. Omar, N. Paunkovic, L. Sheridan, S. Bose, e-print quant-ph/0411065.
\bibitem{lattice}W. D\"{u}r, R. Raussendorf, V. M. Kendon, and H. -J.
Briegel, Phys. Rev. A {\bf66}, 052319 (2002); K. Eckert, J. Mompart,
G. Birkl, M. Lewenstein, Phys. Rev. A {\bf72}, 012327 (2005).
\bibitem{mandel}C. K. Hong, Z. Y. Ou, and L. Mandel,
Phys. Rev. Lett. {\bf59}, 2044 (1987); Z. Y. Ou and L. Mandel, Phys.
Rev. Lett. {\bf61}, 50 (1988); R. Ghosh and L. Mandel, Phys. Rev.
Lett. {\bf59}, 1903 (1987).
\bibitem{kwiat}P. G. Kwiat, K. Mattle, H. Weinfurter, A. Zeilinger, A. V. Sergienko, and Y.
Shih, Phys. Rev. Lett. {\bf75}, 4337 (1995).
\bibitem{nayak}A. Nayak and A. Vishwanath, e-print quant-ph/0010117.
\bibitem{brun}T. A. Brun, H. A. Carteret, and A. Ambainis, Phys. Rev. A {\bf67}, 052317
(2003); Phys. Rev. Lett. {\bf91}, 130602 (2003).
\bibitem{hwp}J. L. O'Brien, G. J. Pryde, A. G. White, T. C. Ralph, D.
Branning, Nature (London){\bf426}, 264 (2003).
\bibitem{bell}J. T. Barreiro, N. K. Langford, N. A. Peters, and P. G.
Kwiat, Phys. Rev. Lett. {\bf95}, 260501 (2005); F. A. Bovino, G.
Castagnoli, A. Ekert, P. Horodecki, C. M. Alves, and A. V.
Sergienko, ibid. {\bf95}, 240407 (2005).
\bibitem{coherent}R. J. Glauber, Phys. Rev. {\bf131}, 2766 (1963).
\bibitem{bornwolf}L. Mandel and E. Wolf, in {\bf Optical Coherence and Quantum Optics}
(Cambridge University Press, 1995) p.128.
\end{thebibliography}
\end{document}